\documentclass[aps,pra,10pt]{revtex4-1}
\usepackage{graphicx,hyperref,jabbrv}

\begin{document}

\title{Optomechanics for quantum technologies}

\author{Shabir Barzanjeh}
\affiliation{Institute for Quantum Science and Technology, University of Calgary, Calgary T2N\,1N4, Alberta, Canada}
\altaffiliation[E-mail address: ]{shabir.barzanjeh@ucalgary.ca}

\author{Andr\'e Xuereb}
\affiliation{Department of Physics, University of Malta, Msida MSD\,2080, Malta}
\altaffiliation[E-mail address: ]{andre.xuereb@um.edu.mt}

\author{Simon Gr\"oblacher}
\affiliation{Kavli Institute of Nanoscience, Department of Quantum Nanoscience, Delft University of Technology, 2628CJ Delft, The Netherlands}

\author{Mauro Paternostro}
\affiliation{Centre for Theoretical Atomic, Molecular, and Optical Physics (CTAMOP), School of Mathematics and Physics, Queen's University, Belfast BT7\,1NN, United Kingdom}

\author{Cindy A Regal}
\affiliation{JILA, University of Colorado and National Institute of Standards and Technology, Boulder CO~80309, USA}
\affiliation{Department of Physics, University of Colorado, Boulder CO~80309, USA}

\author{Eva Weig}
\affiliation{Department of Physics, University of Konstanz, 78457 Konstanz, Germany}
\affiliation{Department of Electrical and Computer Engineering, Technical University Munich, 80333 Munich, Germany}
\affiliation{Munich Center for Quantum Science and Technology (MCQST), Schellingstr.\ 4, 80799 Munich, Germany}

\begin{abstract}
The ability to control the motion of mechanical systems through its interaction with light has opened the door to a plethora of applications in fundamental and applied physics. With experiments routinely reaching the quantum regime, the focus has now turned towards creating and exploiting interesting non-classical states of motion and entanglement in optomechanical systems. Quantumness has also shifted from being the very reason why experiments are constructed to becoming a resource for the investigation of fundamental physics and the creation of quantum technologies. Here, by focusing on opto- and electromechanical platforms we review recent progress in quantum state preparation and entanglement of mechanical systems, together with applications to signal processing and transduction, quantum sensing, topological physics, as well as small-scale thermodynamics.
\end{abstract}

\maketitle

Optomechanics traces its origins to well over a century ago~\cite{1972SciAm.226b..62A}. In 1917, Einstein famously discussed how photons transfer momentum to mirrors~\cite{1917PhyZ...18..121E}; even further back, by 1619 Kepler had hypothesised that the direction in which comet tails point is partly a result of what we would nowadays term radiation pressure~\cite{1619dclt.book.....K}. In optomechanical systems, momentum is exchanged between electromagnetic radiation and mechanical objects. Significant strides forward in manufacturing technology and nano-fabrication made it possible, by the first decade of this century~\cite{2014RvMP...86.1391A}, to start working towards bringing mechanical systems to the quantum regime – close to their quantum ground state, where the quantisation of energy levels and the uncertainty principle significantly affect their dynamics. The potential for optomechanical device fabrication to build on standard microchip technology provides a unique degree of integrability, controllability, and scalability. These fabrication techniques make it possible to couple mechanical resonators to electromagnetic radiation by embedding nano- or micro-scale mechanical resonators in optical cavities~\cite{2011Natur.478...89C, Verhagen2012} or superconducting microwave circuits~\cite{2015Sci...349..952W,2015PhRvL.115x3601P, 2019Natur.570..480B}. This helps to bring optomechanical systems to the fore as candidates for future quantum technologies.

The emergence of quantum technologies relies on the capability of controlling quantum devices and of tailoring their mutual, quantum-coherent, interaction~\cite{2003RSPTA.361.1655D}. Mechanical motion is capable of coupling to a wide variety of natural (e.g., photons) or engineered~\cite{2014QMQM....2....2R,2015PNAS..112.3866K} (e.g., artificial atoms) quantum systems. Moreover, the physical properties of fabricated mechanical devices, in contrast to those of natural quantum systems, can be designed as desired. The ability to bring mechanical systems to the quantum regime and controlling their quantum state paves the way to the building up of a veritable quantum toolbox based on mechanical motion. As discussed by Braginskii and colleagues in the late 1960s~\cite{1977ucp..book.....B}, optomechanical interactions are intimately connected with the quantum-limited measurement~\cite{2010RvMP...82.1155C} of physical phenomena. Beyond being quantum probes useful to measure classical quantities, optomechanical systems are also applicable to measuring genuine quantum phenomena. Recent developments have exploited this ability to create and observe non-classical correlations, including Hanbury-Brown–Twiss correlations~\cite{1956Natur.177...27B} of individual phonons, entanglement~\cite{2009RvMP...81..865H} between two mechanical resonators, and to test Bell nonlocality~\cite{2014RvMP...86..419B} with nano- or micro-scale mechanical resonators.

Here we present an overview of recent progress of optomechanical systems and their technological applications. Our discussion will cover various mechanical systems interacting with electromagnetic radiation in the microwave or optical region of the spectrum; we refer to all of these platforms as \emph{optomechanical systems}. The investigation of non-classical correlations and entanglement between mechanical systems and optical or microwave fields, or between pairs of mechanical systems, is of much interest because it connects quantum physics and tangible objects. Advances in optomechanical systems have motivated their use in applications such as quantum-coherent transducers between optical and microwave fields~\cite{2020QSAT....5b0501L, Lambert}, as quantum sensors for displacement, and to investigate stochastic and quantum thermodynamics~\cite{2009LNP...784.....G,2016ConPh..57..545V}, including as micro- or nanoscopic heat engines and as probes for thermodynamic properties of small-scale systems; this provides one pathway towards autonomous quantum heat engines~\cite{2015NJPh...17k3029B} that exploit, e.g., mechanical quantisation or non-classical correlations. As a final application, we will discuss using optomechanical systems as a platform that employs topological protection for the processing and routing of quantum signals.

\section{The optomechanical interaction}\label{TheoryBox}
A mechanical resonator oscillating at $\omega_{\text{m}}$ coupled to an optical cavity resonant at $\omega_{\text{c}}$ is described by the Hamiltonian~\cite{1995PhRvA..51.2537L,2014RvMP...86.1391A}
\begin{equation}
\hat{H}_{\text{OM}}=\hbar\omega_{\text{m}}\hat{b}^\dagger\hat{b}+\hbar\omega_{\text{c}} \hat{c}^\dagger\hat{c}+\hbar g_0(\hat{b}+\hat{b}^\dagger)\hat{c}^\dagger\hat{c}+\hat H_{\mathrm{d}},
\label{Hamiltonian}
\end{equation}
where $\hat{c}$ and $\hat{b}$ are the optical and mechanical annihilation operators, respectively. The first two terms are the free Hamiltonians of the mechanical and cavity field, the third is the dispersive optomechanical coupling with the single-photon optomechanical coupling strength $g_0$, and $\hat H_\mathrm{d}=\hbar\mathcal{E}(\hat{c} e^{-i\omega_{\text{d}}t}+\text{h.c.})$ represents the driving of the cavity by a coherent electromagnetic field with frequency $\omega_{\text{d}}$ and amplitude $\mathcal{E}$. Typically, the drive is described by the \emph{detuning} $\Delta$, where $\omega _{\text{d}}=\omega_{\text{c}}+\Delta$; conventionally, $\Delta<0$ ($\Delta>0$) is termed \emph{red} (\emph{blue}) detuning. In a frame rotating with the driving field, Eq.~(\ref{Hamiltonian}) transforms to
\begin{equation}
\hat{H}_{\text{OM}}^\prime=\hbar\omega_{\text{m}}\hat{b}^{\dagger}\hat{b}+ \hbar\bigl[-\Delta+g_{0}(\hat{b}+\hat{b}^{\dagger})\bigr]\hat{c}^{\dagger}\hat{c}+\hat{H}_{\text{d}}^\prime,
\label{HamiltonianIntFrame}
\end{equation}
where the Hamiltonian associated with the driving field becomes $\hat{H}_{\text{d}}^\prime=\hbar\mathcal{E}(\hat{c}+\hat{c}^{\dagger})$. This Hamiltonian is cubic in the field operators and predicts a rich dynamics~\cite{2011PhRvL.107f3602N,2011PhRvL.107f3601R}. In many experiments the coupling strength $g_0$ is significantly smaller than the other frequencies involved, leading to a dynamics that can be well-approximated by a linearised~\cite{2014RvMP...86.1391A} Hamiltonian.

We linearise Eq.~(\ref{HamiltonianIntFrame}) by expanding the optical mode around its steady-state amplitude, $\hat{a}=\hat{c}-\alpha$, where $\vert\alpha\vert^2=4\vert\mathcal{E}\vert^{2}\big/(\kappa^{2}+4\Delta^{2})$ is the mean number of intra-cavity photons and $\kappa$ is the cavity decay rate. The linearised Hamiltonian transforms to
\begin{equation}
\hat{H}_{\text{lin}}=-\hbar\Delta\hat{a}^{\dagger}\hat{a}+\hbar\omega_{\text{m}}\hat{b}^{\dagger}\hat{b}+\hbar G(\hat{a}+\hat{a}^{\dagger})(\hat{b}+\hat{b}^{\dagger}),
\label{HamiltonianLinearised}
\end{equation}
where $G=g_{0}\vert\alpha\vert$ is the optomechanical coupling strength; the phase of $\alpha$ does not play a role in the dynamics and can be neglected, but its amplitude acts to amplify the coupling strength. Moving to a frame rotating at the optical and mechanical frequencies, i.e., moving to an interaction picture with respect to $-\hbar\Delta\hat{a}^{\dagger}\hat{a}+\hbar\omega_{\text{m}}\hat{b}^{\dagger}\hat{b}$, Hamiltonian~(\ref{HamiltonianLinearised}) becomes
\begin{equation}
\hat{H}_{\text{lin}}^\prime=\hbar G(\hat{a}e^{\mathrm{i}\Delta t}+\hat{a}^{\dagger}e^{-\mathrm{i}\Delta t})(\hat{b}e^{-\mathrm{i}\omega _{m}t}+\hat{b}^{\dagger}e^{\mathrm{i}\omega_{\text{m}}t}).
\label{HamiltonianLinearisedIntFrame}
\end{equation}

\begin{figure*}[th!]
	\centering{\includegraphics[width=0.55\textwidth]{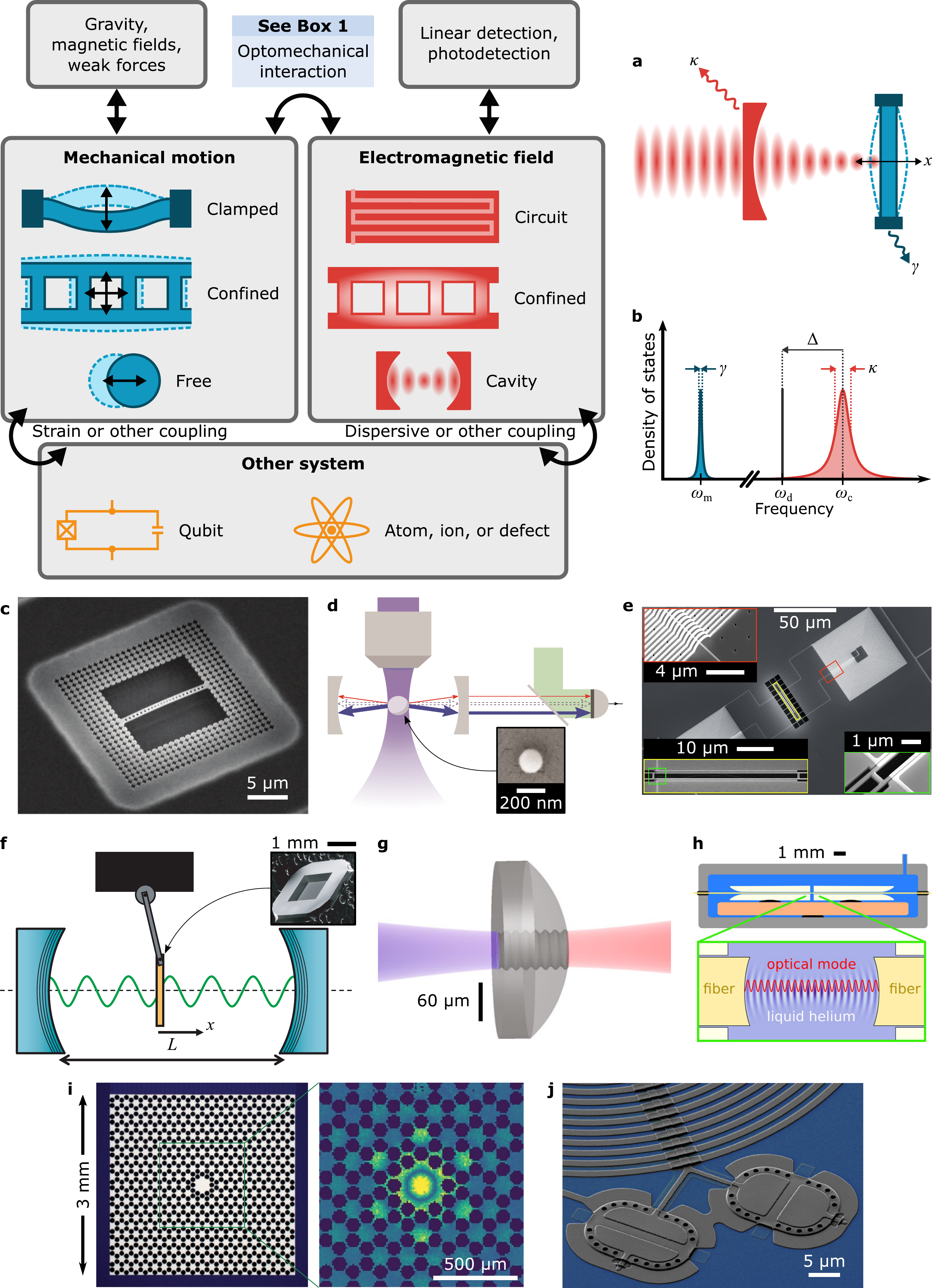}}
	\caption{\textbf{Platforms for studying optomechanical science.} (Top left) The optomechanical interaction and a range of couplings enabled by mechanical, electrical, and optical functionality. Optomechanical devices can interact with qubits, atoms, or other systems, for instance by means of coupling to strain variations, in the case of defects embedded in mechanical resonators. Optomechanical systems can be used for measuring weak forces (e.g., gravity), faint signals (electric or magnetic fields), or small masses. Nonlinear resources can be introduced in optomechanical systems through single-photon counting. \textbf{a},~In the prototypical model, an optical cavity is coupled to a mechanical resonator through its displacement $x$. The physical mechanism underlying this interaction is the change in the resonance frequency of the cavity induced by a displacement of the resonator. We denote the mechanical and optical loss rates by $\gamma$ and $\kappa$, respectively. \textbf{b},~Density of states of the mechanical and optical resonances, having resonance frequencies $\omega_\mathrm{m}$ and $\omega_\mathrm{c}$, respectively, and driven by a coherent tone whose frequency is $\omega_\mathrm{d}$; in this illustration we show the red-detuned regime where the detuning is negative, $\Delta=\omega_\mathrm{d}-\omega_\mathrm{c}<0$. Optomechanical phenomena, e.g., sideband cooling~\cite{2007PhRvL..99i3902M}, can often be understood physically by considering the relative locations of the driving fields and the peaks in the density of states. \textbf{c},~A phononic- and photonic-crystal cavity~\cite{2011Natur.478...89C} with a mechanical resonator in the center suspended from a phononic shield that assists in minimising the effects of the environment. \textbf{d},~A dielectric nano-sphere~\cite{2020Sci...367..892D}, shown in the inset, that can be optically levitated and cooled in an optical cavity. \textbf{e},~A mechanical resonator~\cite{2019Natur.570..480B} (lower left and right insets) coupled to a gigahertz microwave circuit (upper left). \textbf{f},~A vibrating membrane~\cite{2008Natur.452...72T}, shown in the inset, coupled dispersively to a high-finesse Fabry–Pérot cavity of fixed length $L$. \textbf{g},~Brillouin optomechanics, which exploits bulk crystal vibrations~\cite{2018NatPh..14..601R}. \textbf{h},~Optomechanical physics in superfluid helium~\cite{2019PhRvL.122o3601S} with an optical cavity bounded by two optical fibres and a mechanical resonator formed from an acoustic mode in the helium. \textbf{i},~An optomechanical membrane incorporating a phononic crystal that possesses very long mechanical coherence times~\cite{Tsaturyan2017}; a membrane of this kind could be used inside a cavity much like the one in panel \textbf{f}. \textbf{j},~A pair of parallel-plate capacitors in an $LC$ resonator whose capacitance is modified by the vibrational motion of their plates. Figure adapted with permission from:\ panel \textbf{c}, ref.~\cite{2011Natur.478...89C}, Springer Nature Ltd; panel \textbf{d}, ref.~\cite{2020Sci...367..892D}, AAAS; panel \textbf{e}, ref.~\cite{2019Natur.570..480B}, Springer Nature Ltd.; panel \textbf{f}, ref.~\cite{2018NatPh..14..601R}, Springer Nature Ltd.; panel \textbf{g}, ref.~\cite{2008Natur.452...72T}, Springer Nature Ltd.; panel \textbf{h}, ref.~\cite{2019PhRvL.122o3601S}, APS; panel \textbf{i}, ref.~\cite{Tsaturyan2017}, Springer Nature Ltd.; panel \textbf{j}, ref.~\cite{Kotler622}, AAAS.}
\label{Platforms}
\end{figure*}

By properly setting the detuning, the optical and mechanical systems can be made to interact in different ways. If the driving frequency is chosen so that $\Delta=-\omega_{\text{m}}$ and the system is in the \emph{sideband-resolved regime}, $\omega_{\text{m}}\gg\kappa$, it is possible to use a rotating-wave approximation to drop the rapidly rotating terms oscillating at $\pm2\omega_{\text{m}}$. This results in an interaction reminiscent of a beam splitter, $\hat{H}_{\text{BS}}=\hbar G(\hat{a}\hat b^{\dagger}+\hat{a}^{\dagger}\hat{b})$, where excitations are exchanged between photons and phonons at rate $G$. This type of interaction can be used to cool down the mechanical motion to its ground state through \emph{sideband cooling}, where the mechanical resonator scatters radiation preferentially into resonance with the cavity mode, which lies at a higher frequency than the coherent field that is driving the cavity, Fig.~\ref{Platforms}(b). Setting the detuning to $\Delta=\omega_{\text{m}}$ results in a parametric down-conversion type interaction, $\hat{H}_{\text{PDC}}=\hbar G(\hat{a}^{\dagger}\hat{b}^{\dagger}+\hat{a}\hat{b})$, which generates entanglement between photons and phonons.

\section{Mechanical systems in the quantum regime}
Our discussion in this review will focus mainly on optomechanical devices where a mechanical resonator is coupled to a resonant electromagnetic cavity. Although it is by no means a limitation inherent in the physics of optomechanical systems~\cite{2008PhRvL.100x0801K,2009PhRvA..79e3810X}, the strength of the radiation pressure interaction is typically much greater in cavity optomechanical systems than in free-space devices, making them more technologically relevant. Section~\ref{TheoryBox} gives a brief overview of the theoretical description of a mechanical object interacting with electromagnetic radiation in a cavity~\cite{2014RvMP...86.1391A}; a variety of platforms that have been used to demonstrate or exploit optomechanical physics are illustrated in Fig.~\ref{Platforms}. In this section, we start by reviewing progress in cooling mechanical resonators to their motional ground state before discussing the preparation of non-classical states.

\subsection{Ground state cooling.}
The preparation of mechanical harmonic oscillators in the ground (i.e., lowest-energy, or vacuum) state of motion was a driving force in optomechanics for nearly a decade~\cite{2014RvMP...86.1391A}. The aim of this quest was to overcome the effects of thermal fluctuations that long precluded study of quantum phenomena in nanomechanical devices. Success in a wide variety of platforms has firmly placed mechanical systems in the quantum regime; this has now become a routine starting point for a variety of complex quantum experiments with mechanical degrees of freedom. High-frequency gigahertz mechanical resonators can be cooled directly to their ground state in a dilution refrigerator. The first experiment to report the use of refrigeration to cool a mechanical resonator to its ground state~\cite{2010Natur.464..697O} used a superconducting quantum bit (qubit) to read out the mechanical motion of the resonator. This ushered in the study of mechanical elements coupled to qubits in quantum acoustics experiments~\cite{2017Sci...358..199C,2020NatPh..16..257C}.

Following this demonstration of reaching the ground state, sideband cooling techniques (see Sec.~\ref{TheoryBox}) were used to place the mechanical resonators in electro-~\cite{2011Natur.475..359T} and optomechanical systems~\cite{2011Natur.478...89C} in their respective ground states. This technique has made the study of resonators whose frequency is as low as megahertz~\cite{peterson2016laser} or below in their quantum ground state possible; alternative cooling techniques make use of active feedback~\cite{2018Natur.563...53R} or squeezed light~\cite{2017Natur.541..191C}. An attractive feature of optomechanical systems is that the ratio of the strengths of the sidebands that the motion imprints on the outgoing electromagnetic field can be used to determine the absolute temperature of their state of motion~\cite{2012PhRvL.108c3602S}.

While nanofabricated solids were the first mechanical objects to be brought to the ground state, recent years have seen expansion of the platforms for which quantum control is possible, including optically levitated sub-megahertz mechanical resonators~\cite{2020ConPh..61..155M}, which can be brought into their ground state from room temperature~\cite{2020Sci...367..892D} or without the use of an optical cavity~\cite{Magrini2021,Tebbenjohanns2021}; optomechanics in superfluid helium coupled to a cavity~\cite{2019PhRvL.122o3601S,2020NatPh..16..417H}; and large mass bulk acoustic waves, whose substantial cooling is enabled by large Brillouin couplings~\cite{2018NatPh..14..601R}.

\begin{figure}[t!]
	\centering{\includegraphics[width=0.4\columnwidth]{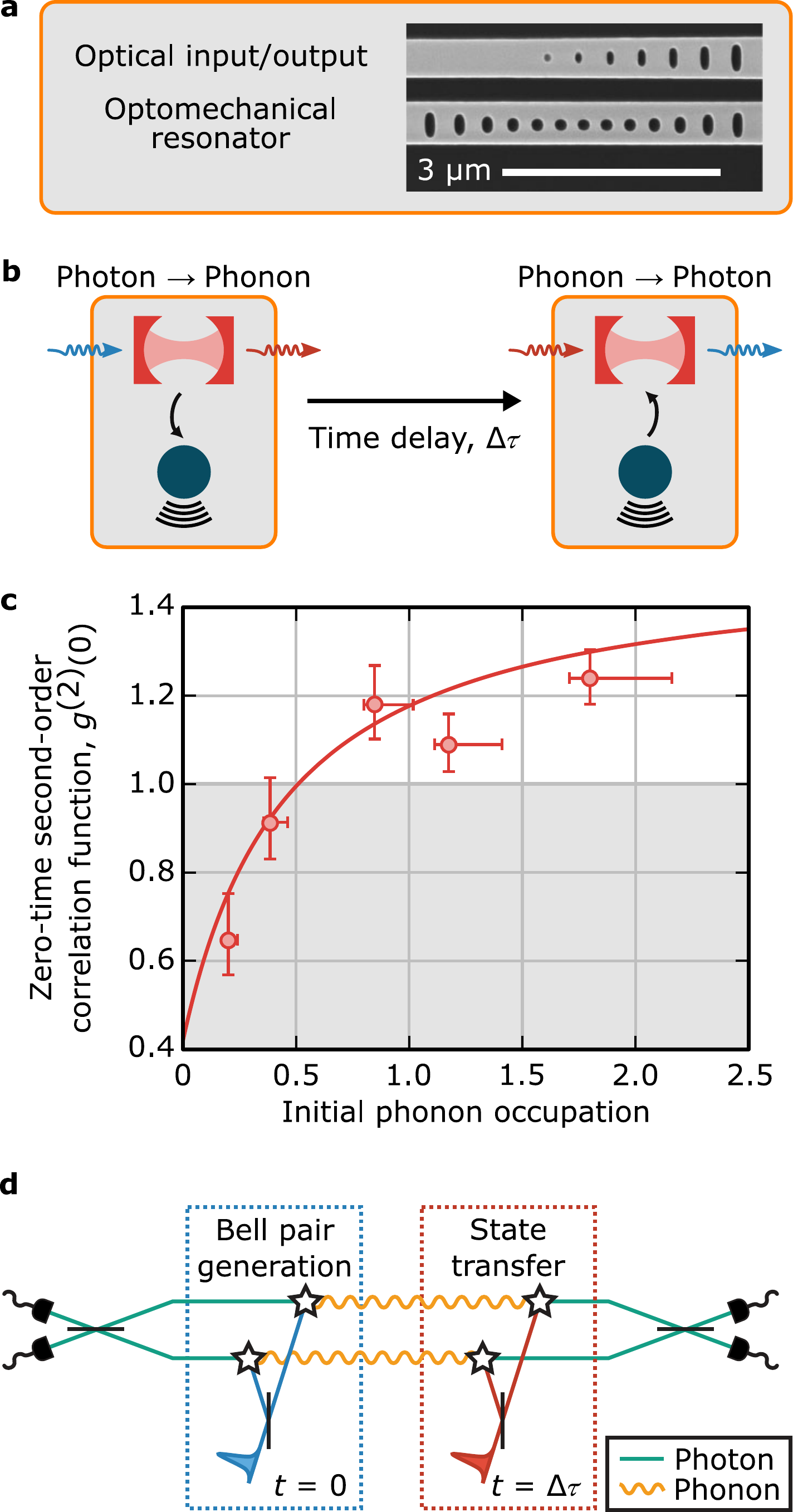}}
	\caption{\textbf{Non-classical correlations in and between mechanical resonators.} \textbf{a},~A scanning electron microscope image of an optomechanical device~\cite{2017Sci...358..203H} used to generate correlations between phonons. \textbf{b},~The experimental protocol employed consists of optically exciting a mechanical resonator with a single phonon and reading it back out after a time delay $\Delta \tau$. \textbf{c},~The zero-time second-order correlation function $g^{(2)}(0)$; values below $1$ denote non-classical Hanbury-Brown–Twiss correlations between phonons~\cite{2017Sci...358..203H}. \textbf{d},~Experimental protocol used to demonstrate a violation of a Bell inequality in an optomechanical system with two mechanical resonators~\cite{2018PhRvL.121v0404M}; the state is initialised at $t=0$ and read out optically after a time delay $t=\Delta \tau$, with entanglement between the mechanical resonators converted to correlations between photons. Each branch consists of a device similar to that in panel \textbf{a.} Figure adapted with permission from:\ panels \textbf{a} and \textbf{c}, ref.~\cite{2017Sci...358..203H}, AAAS; panel \textbf{d}, ref.~\cite{2018PhRvL.121v0404M}, APS.}
\label{Correlations}
\end{figure}

\subsection{Quantum state preparation of a mechanical resonator.}
The preparation of non-classical states in optomechanical systems can target two distinct families of states. On the one hand, the dynamics arising from the linearised optomechanical Hamiltonian~(\ref{HamiltonianLinearised}) is limited to the creation of Gaussian states when starting from a thermal state. While manifestly non-classical in a variety of ways, and in fact capable of demonstrating entanglement, Gaussian states can be described by means of a non-negative Wigner quasiprobability distribution~\cite{fa994293430a4d549879f0736c0c567a} and thus behave somewhat similarly to classical states. On the other hand the full nonlinear optomechanical Hamiltonian becomes significant only when the single-photon coupling strength is large. Coupling to a qubit~\cite{2017NatPh..13.1163R}, or performing strong measurements~\cite{Clerk2020}, can be used to prepare non-Gaussian states of motion, such as Fock states or Schr\"odinger cat states~\cite{2005qoai.book.....V}.

A restricted, albeit very important and well-studied, class of non-classical states is that of squeezed states, which are intimately tied to a cornerstone of quantum physics, the Heisenberg uncertainty principle. This principle sets the scale of the zero-point fluctuations of a mechanical resonator in its quantum ground state. The optomechanical interaction with combined blue- and red-detuned driving tones (see Sec.~\ref{TheoryBox}) allows to suppress one of the quadratures of the mechanical fluctuations below the zero point level, at the expense of increasing the fluctuations in the orthogonal quadrature. The first experimental demonstrations of optomechanical squeezing was accomplished in electromechanical systems featuring a mechanically compliant, superconducting drum ~\cite{2015Sci...349..952W,2015PhRvL.115x3601P,2015PhRvX...5d1037L} or nanostring~\cite{2017NatCo...8..953B, 2019Natur.570..480B} parametrically coupled to a microwave cavity like the one shown in Fig.~\ref{Platforms}(e). Evidence of squeezed states of motion of a mechanical resonator is most directly obtained from an analysis of the motional quadrature spectra~\cite{2019PhRvL.123r3603D}.

More recently, the strong piezoelectric coupling of superconducting qubits to mechanical resonators~\cite{2010Natur.464..697O} has grown into the field of quantum acoustics. Systems that strongly couple qubits to mechanical motion enable the counting of phonons in mechanical resonators, placing this motion in highly nonclassical states such as multi-phonon Fock states in an electromechanical system~\cite{2018Natur.563..666C}, and coupling propagating phonons to qubits ~\cite{2017NatCo...8..975M,2018PhRvL.120v7701M,2020PhRvX..10b1055B}. In this context, quantum acoustics has found applications in phonon creation and detection schemes~\cite{2018PhRvL.121r3601V,2019Natur.571..537A,2019PhRvX...9b1056S}. For a recent review of the prospects of qubit--mechanical coupled systems we refer the reader to ref.~\cite{2020NatPh..16..257C}.

\subsection{Entanglement and non-classical correlations.}
One of the hallmarks of quantum physics is the notion of correlations stronger than allowed by classical mechanics, most famously quantum entanglement~\cite{2009RvMP...81..865H}. In quantum mechanics, if two systems are entangled, their individual states before a measurement cannot in principle be described independently; they have to be treated as a single system, no matter how different or far apart they are from one another. A measurement on one seems to directly and instantaneously influence the quantum state of the other. Demonstrating these quantum correlations was a long sought-after experimental challenge in optomechanics, with the aim of unequivocally demonstrating the existence of a mechanical system in a non-classical state of motion.

\begin{figure}[t!]
	\centering{\includegraphics[width=0.4\columnwidth]{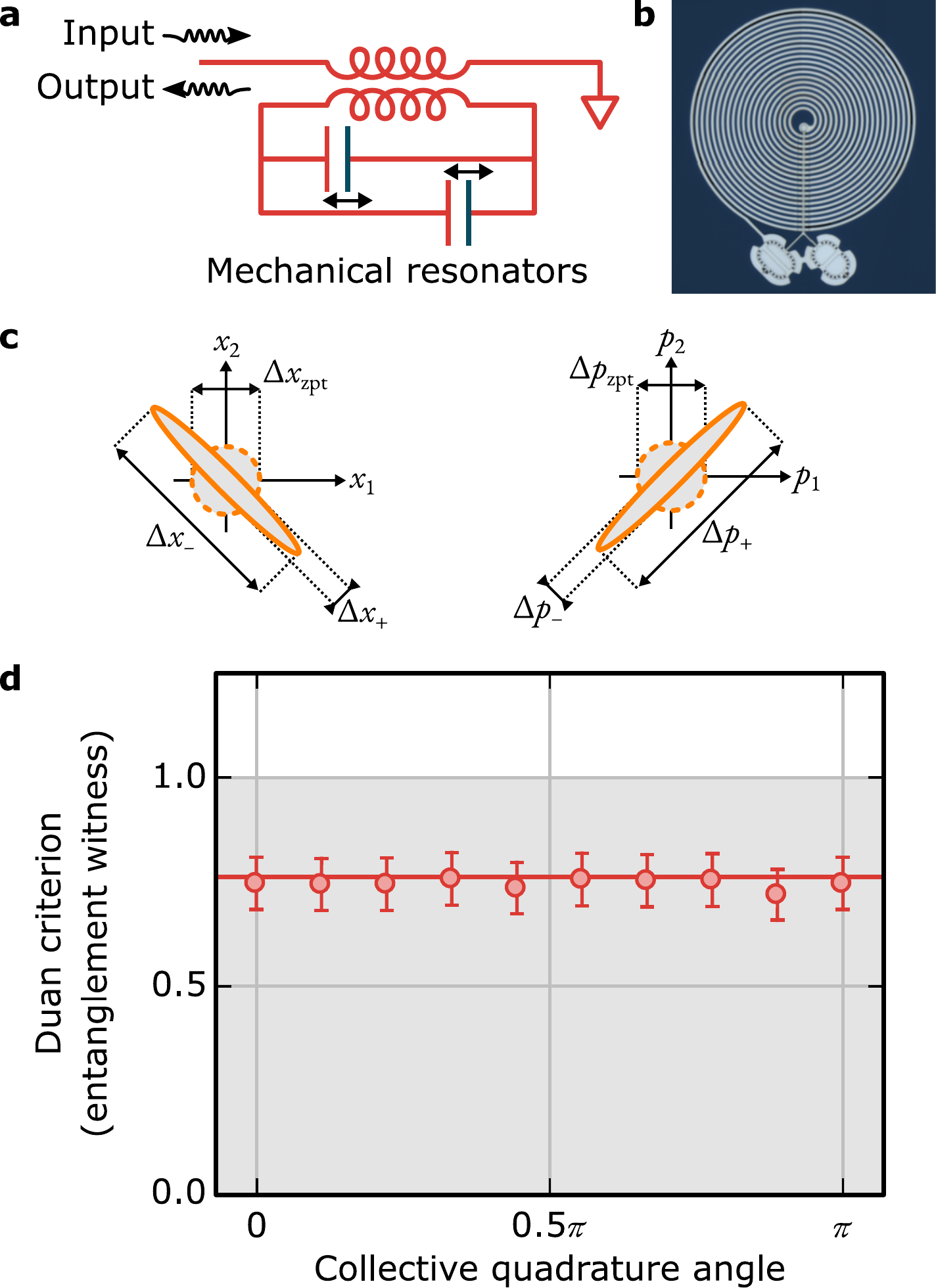}}
	\caption{\textbf{Entanglement of massive mechanical resonators.} Shown here are \textbf{a},~the schematic of the experiment, consisting of two mechanical resonators coupled to a single electromagnetic cavity, in this case an $LC$ resonator that is coupled inductively to an input and output port; and \textbf{b},~a micrograph of the device used to conduct an experiment in which a superconducting microwave resonator is coupled to two mechanical drum-head resonators. \textbf{c}, Entanglement between two mechanical resonators is best understood in terms of uncertainties. When the resonators are in an entangled state, the uncertainties in two collective coordinates (${\Delta}x_-$, the uncertainty in the difference of the positions, and ${\Delta}p_+$, the uncertainty in the sum of the momenta) go below the size of the zero-point fluctuations (${\Delta}x_{\text{zpt}}$, ${\Delta}p_{\text{zpt}}$). \textbf{d},~The Simon–Duan criterion $({\Delta}x_-/{\Delta}x_{\text{zpt}}+{\Delta}p_+/{\Delta}p_{\text{zpt}})/2$, which is a witness for the presence of entanglement when its value dips below unity; the horizontal axis corresponds to angle of the collective mechanical quadrature along which the entanglement was produced, demonstrating the capability to generate entanglement along any chosen collective quadrature. The solid red horizontal line is the theoretical prediction. Figure adapted with permission from: panel \textbf{b}, ref.~\cite{Kotler622}, AAAS; panel \textbf{c}, ref.~\cite{2018Natur.556..478O}, Springer Nature Ltd.; panel \textbf{d}, ref.~\cite{Mercier}, AAAS.}
\label{Multimode}
\end{figure}

The first experimental demonstration of an entangled state was realised in an electromechanical system where the mechanical motion of a superconducting drum was entangled with a travelling microwave field~\cite{2013Sci...342..710P}. In the optical regime, the first experiment realising non-classical correlations between the light field and mechanics was performed using a gigahertz photonic- and phononic-crystal cavity~\cite{2016Natur.530..313R}. More recently, continuous-variable entanglement generation of microwave radiation using a mechanical resonator was also demonstrated~\cite{2019Natur.570..480B}.

\begin{figure}[th!]
	\centering{\includegraphics[width=0.4\columnwidth]{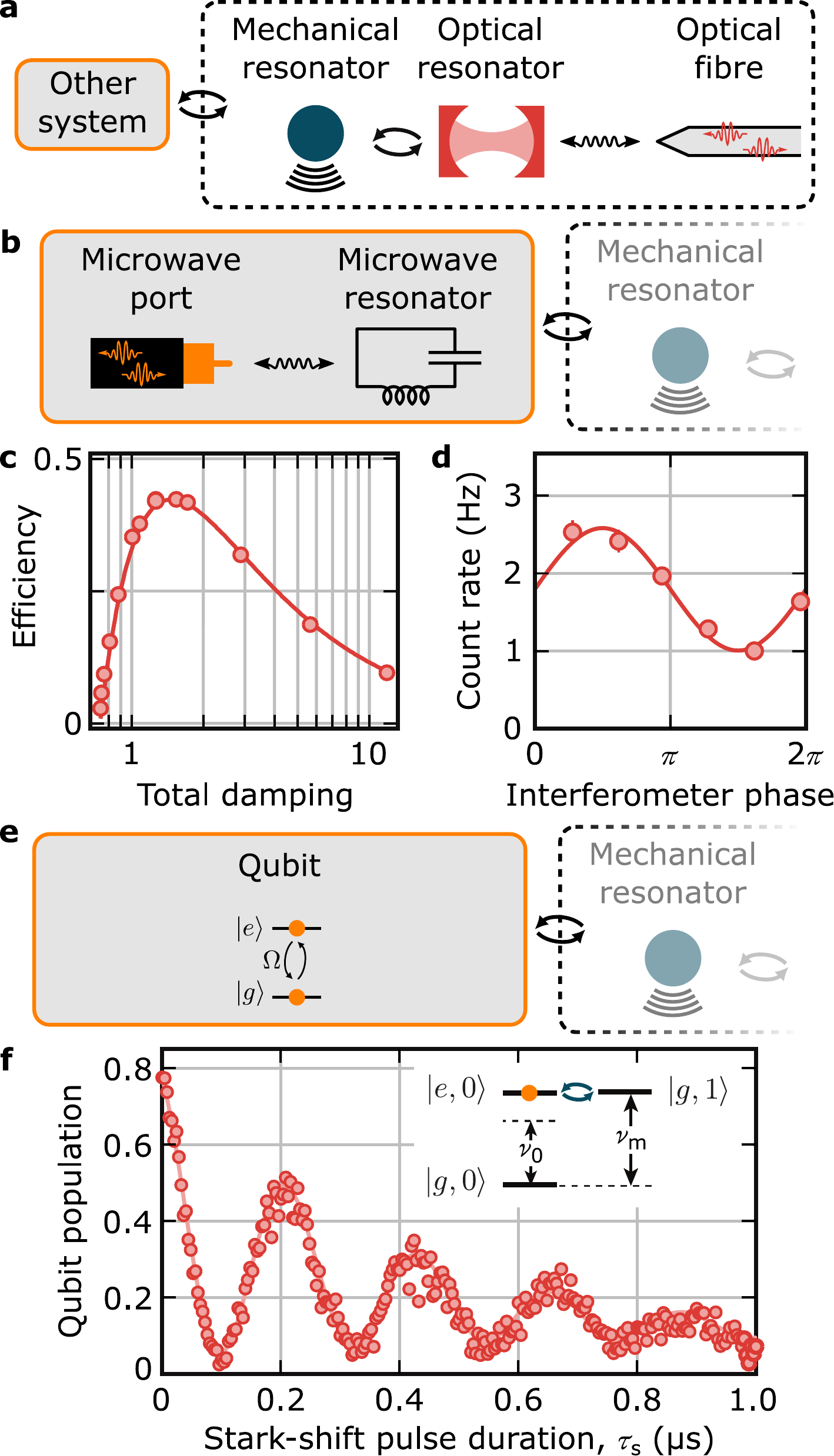}}
	\caption{\textbf{Efficient and coherent transduction in optomechanical systems.} \textbf{a},~Principle of operation of an optomechanical transducer~\cite{2018NatPh..14.1038H}. \textbf{b},~Schematic of an optomechanical transducer coupled to a microwave resonator, forming a microwave-to-optical photon converter. \textbf{c},~The efficiency of this converter; its maximum efficiency was measured to be approximately 47\%~\cite{2018NatPh..14.1038H}, obtained by setting the mechanical damping rate appropriately. \textbf{d},~Coherent conversion between optics and microwaves using a mechanical transducer~\cite{2019NatPh..16...69F}; interference of the original microwave signal with a phase-delayed signal converted to light and back is evidence of phase-coherent transduction between optical and microwave radiation. \textbf{e},~Principle of operation of an optomechanical transducer that is used to couple a qubit to an optical fibre~\cite{Mirhosseini2020}. \textbf{f},~Quantum-coherent inter-conversion between microwave photons and phonons in an electromechanical circuit~\cite{Mirhosseini2020}. In this experiment, Rabi oscillations shift excitations coherently between the qubit and the mechanical resonator; this is evidenced by the oscillating qubit population. The inset shows the level structure for the coupled qubit (levels $g$ and $e$) and mechanical oscillator (number states $0$ and $1$). Figure adapted with permission from:\ panels \textbf{a}, \textbf{e}, and \textbf{f}, ref.~\cite{Mirhosseini2020}, Springer Nature Ltd.; panel \textbf{c}, ref.~\cite{2018NatPh..14.1038H}, Springer Nature Ltd.; panel \textbf{d} ref.~\cite{2019NatPh..16...69F}, Springer Nature Ltd.}
\label{Conversion}
\end{figure}

The next conceptual step after creating non-classical correlations between light and mechanics is quantum correlations between two mechanical systems. A photonic- and phononic-crystal cavity was also used to perform phonon counting ~\cite{2015Natur.520..522C} or to demonstrate the well-known Hanbury-Brown–Twiss correlations between phonons~\cite{2017Sci...358..203H}, Fig.~\ref{Correlations}(a–c). The effect of quantum correlations is stronger the closer the mechanical resonator is to its ground state; this is seen in Fig.~\ref{Correlations}(c), where the correlation function $g^{(2)}(0)$ dips below $1$, which is forbidden for classical excitations, when the initial temperature is sufficiently low. The violation of a Bell inequality is seen as the ultimate test of non-classicality, as it allows to probe non-classical correlations independently of the validity of quantum physics and -- under suitable conditions -- in a device-independent manner. For an entangled optomechanical state, a Bell violation was also first shown using high-frequency photonic- and phononic-crystal cavities~\cite{2018PhRvL.121v0404M}, Fig.~\ref{Correlations}(d). Creating a non-classical state shared between two mechanical oscillators that are physically separated from one another was also experimentally realised optically using phononic-crystal cavities~\cite{2018Natur.556..473R} and simultaneously in the microwave regime with two superconducting drum resonators~\cite{2018Natur.556..478O, Mercier, Kotler622}, Fig.~\ref{Multimode}(a–b). By calculating an entanglement witness, such as the Simon–Duan criterion~\cite{2009RvMP...81..865H} that is illustrated in Fig.~\ref{Multimode}(d), one can verify the presence of entanglement between mechanical resonators. It is worth mentioning in this context that entanglement has also been observed between mechanical and spin systems separated by a distance of $1$\,m~\cite{2021NatPh..17..228T}.

Most of these experiments were aimed at showing the general feasibility of creating non-classical correlations in an optomechanical system. However rapid improvements, for example directly reading out states of mechanical motion~\cite{Kotler622}, should allow the use of entanglement as a starting point for new fundamental experiments and as a resource for novel quantum technologies, including quantum repeaters~\cite{2021arXiv210402080F}, and sensors~\cite{2015PhRvL.114h0503B, 2020SciA....6B.451B}.

\section{An optomechanical toolbox}
We now turn our attention to some key applications in emerging quantum technologies of mechanical systems. We will discuss the processing of signals, where significant progress has been made recently in the coherent inter-conversion between optical and microwave frequencies, and in the optomechanical sensing of displacement. We also focus on how stochastic and quantum thermodynamics, which has also seen an explosion of interest in recent years~\cite{2019qtit.book.....D}, can be explored by using optomechanical systems either as heat engines or as probes of thermodynamic quantities or processes, and finish off by reviewing recent progress in designing topologically-protected optical or mechanical signal processing devices.

\subsection{Transduction and signal processing.}
A key area where mechanical devices have shown unique promise is in signal transduction in contexts where disparate resources are to be connected~\cite{2010PhRvA..82e3806T,2012PhRvL.109m0503B,2020QSAT....5b0501L}, or in cases where the properties of phonons have advantages over photons~\cite{2012NJPh...14k5004H, 2017PhRvL.118m3601V, 2018PhRvL.121d0501P}. This is particularly apparent in electrical quantum systems~\cite{2020NatPh..16..257C} that can be connected to optical photons by means of optomechanical devices. For emerging quantum technologies there is a pressing need for devices that can efficiently and noiselessly convert signals from microwave to optical frequencies, e.g., in order to interface computational devices based on superconducting qubits with telecommunications networks. Mechanical motion has been shown to simultaneously couple well to both optical and electrical signals. Despite a plethora of other potential technologies, such as direct electro-optic conversion in nonlinear crystals like lithium niobate~\cite{2016Optic...3..597R,2019npjQI...5..108R,2020Optic...7.1737M, PRXQuantum.1.020315,2020NatCo..11.3237H}, mechanical systems are thus far the only platform to have experimentally demonstrated the most important characteristics required of such a transduction device, including conversion that approaches unit efficiency~\cite{2014NatPh..10..321A}, low-noise operation~\cite{2019NatPh..16...69F}, and conversion of single-photon-level signals by coupling to a superconducting qubit~\cite{Mirhosseini2020}.

Figure~\ref{Conversion} illustrates different concepts for transduction and phase-coherent conversion between microwave signals from a superconducting qubit to optical light. Gigahertz-frequency piezo-electric devices can be used to couple directly to both qubits and light through a photonic crystal cavity. In these experiments, Fig.~\ref{Conversion}(a–c), a mechanical object such as a thin membrane is simultaneously coupled to an optical and to a microwave cavity, as shown schematically in Fig.~\ref{Conversion}(b), resulting in the standard optomechanical interaction linearised by a strong pump. This kind of double parametric coupling has the advantage of not requiring frequency matching or piezoelectric materials. Well-understood signatures of coherent conversion, visible interferometric fringes and Rabi oscillations, have been observed and are shown in Fig.~\ref{Conversion}(d–f). Early experiments showed microwave driving of this kind of system can yield an optical signal~\cite{2013NatPh...9..712B}, and recent developments of this concept have observed optical photons transduced from a superconducting qubit~\cite{2019NatPh..16...69F,Mirhosseini2020}. Nanoscale devices, however currently have efficiencies far from what is needed for heterogeneous quantum networks. Approaches using low-frequency mechanical resonators have achieved much higher conversion efficiencies close to $50$\%~\cite{2014NatPh..10..321A,2018NatPh..14.1038H,2020NatCo..11.4460A}, yet still with some added noise that must be removed for quantum operation.

\subsection{Displacement sensing with optomechanics.}
Optomechanics is also inextricably linked to precise interferometric measurement of mechanical motion. In fact early thinking on optical radiation pressure effects in mechanical devices stemmed from explorations into fundamental limits of precision sensors~\cite{1977ucp..book.....B,caves1980measurement}. Small scale optomechanical devices have played a role in understanding application of quantum sensing concepts to detecting mechanical motion. Shot noise places a limit on the imprecision of any interferometer, and while the ratio of signal to noise can be improved by increasing probe strength, eventually quantum backaction in the form of the shot noise of radiation pressure will obscure the measurement. The balance of these two sources in a typical interferometer configuration is often referred to as the standard quantum limit (SQL). Seminal works put forth the idea of backaction-evading measurements~\cite{braginsky1980quantum}, which have been applied for example in microwave cavity-based devices~\cite{hertzberg2010back,ockeloen2016quantum}. An active area of research is constructing \emph{quantum mechanics free} subspaces where all variables are quantum nondemolition observables~\cite{tsang2012evading,de2021quantum,2021NatPh..17..228T}.

Squeezing of the quantum noise of light is a key opportunity for improving mechanical detectors. The use of squeezed light to reduce imprecision due to shot noise has been demonstrated in pioneering experiments and even recently in the LIGO gravitational-wave observatories~\cite{aasi2013enhanced,tse2019quantum}. The use of squeezed resources becomes even more effective when radiation pressure shot noise plays a role in the observed quantum limits~\cite{purdy2013observation,Cripe2019}. In the regime where radiation pressure shot noise is a dominant effect on the motion of mechanical devices, as is natural in optomechanical devices, so-called ponderomotive squeezing of light can be obtained by harnessing the entangling interaction that occurs with blue-detuned light in cavity optomechanics (see Sec.~\ref{TheoryBox})~\cite{brooks2012non,2013Natur.500..185S,2013PhRvX...3c1012P,clark2016}. This illustrates an interesting feature of optomechanics; namely, while light can be used to manipulate mechanical motion in the quantum regime, mechanical motion, even as a classical object, can also manipulate the quantum properties of light. Ponderomotive squeezing can be harnessed to improve displacement detection~\cite{vyatchanin1995quantum}, a concept that has been shown experimentally near the standard quantum limit of an interferometer and beyond~\cite{Kampel2017improving,mason2019continuous}, and has now even been observed in the kilogram-scale mirrors of LIGO~\cite{2018JLwT...36.3919L}. The quantum sensing frontier is ever-expanding in gravitational-wave detectors, and a next goal is the injection of frequency-dependent squeezing to achieve broadband gains in sensitivity~\cite{Yap2020,mcculler2020frequency}.

\subsection{Thermodynamics and thermal machines.}
The prototypical optomechanical setting, Fig.~\ref{Platforms}(a), bears strong resemblance with the paradigm of a thermodynamic system, namely a chamber filled with a gas and endowed with a movable piston. This analogy has been recently pushed all the way to the proposal and implementation of prototype heat engines based on optomechanical dynamics with the ambition to explore finite-time thermodynamics in both classical and quantum contexts.

In practice, the Hamiltonian~(\ref{HamiltonianLinearised}) is complemented by the interaction of the mechanical motion and the electromagnetic field with their respective environments which can be at different temperatures. The optomechanical coupling between the cavity field and mechanical motion results in hybrid photon--phonon excitations that can be tuned from phonon-like to photon-like by varying the driving laser frequency. In turn, this implies that the corresponding excitations are coupled to either the phononic or the photonic environments, respectively. The different effective temperatures of these two thermal reservoirs have been exploited to arrange for an Otto cycle~\cite{2014PhRvL.112o0602Z,PhysRevA.90.023819,PhysRevA.92.033854}. Feedback-based mechanisms can be put in place to bypass the necessity of strong optomechanical coupling~\cite{Abari_2019} that would otherwise be required to significantly hybridise the photonic and phononic excitations. While in this scheme a thermodynamic cycle is induced by modulating a laser frequency, alternative approaches to continuous work extraction make use of a temperature gradient between the thermal baths between which the engine operates to sustain mechanical oscillations~\cite{Mari_2015} in an autonomous manner, or a broadband heat bath with no coherence or phase-locking~\cite{Kurizki_2015}. A particularly interesting platform for work extraction is the levitated optomechanical scenario consisting of a sub-micron particle optically trapped in a harmonic potential with controllable mechanical frequency~\cite{2015PhRvL.114r3602D}, Fig.~\ref{Platforms}(d). The thermal environment embodied by the residual gas in the trap provides the heat bath for this system. A Stirling engine can then be operated by weakly coupling the under-damped mechanical motion to the optical cavity.

The potential of optomechanical systems for sensing and probing extends to the assessment of crucial quantities in finite-time non-equilibrium thermodynamics. Specifically, it offers the possibility to study the interplay between interaction and incoherent dynamics induced by the coupling with environments. This interplay enables the preparation of non-equilibrium steady states and the study of the entropy production process resulting from their deviation from thermal equilibrium~\cite{2020arXiv200907668L}. In optomechanical systems, entropy production -- which in essence is a non-equilibrium formulation of the second principle of thermodynamics~\cite{2020arXiv200907668L} -- can be experimentally characterised by accessing the steady-state noise spectral density~\cite{2018PhRvL.121p0604B}. This quantity provides key information on the entropy fluxes from the optical and mechanical subsystems to the respective thermal environment. This idea has been extended to address the entropic term resulting from the continuous measurement to which the optomechanical system is exposed, thus demonstrating the substantial contribution arising from the \emph{information} gathered through measurement~\cite{2019arXiv190809382B}. The entropy production resulting from the relaxation of the non-equilibrium initial state of motion of a mechanical system, and the characterisation of the environment into which the system is embedded~\cite{2014NatNa...9..425M}, have been at the focus of significant recent experimental endeavours.

\begin{figure}[t!]
	\centering{\includegraphics[width=0.4\columnwidth]{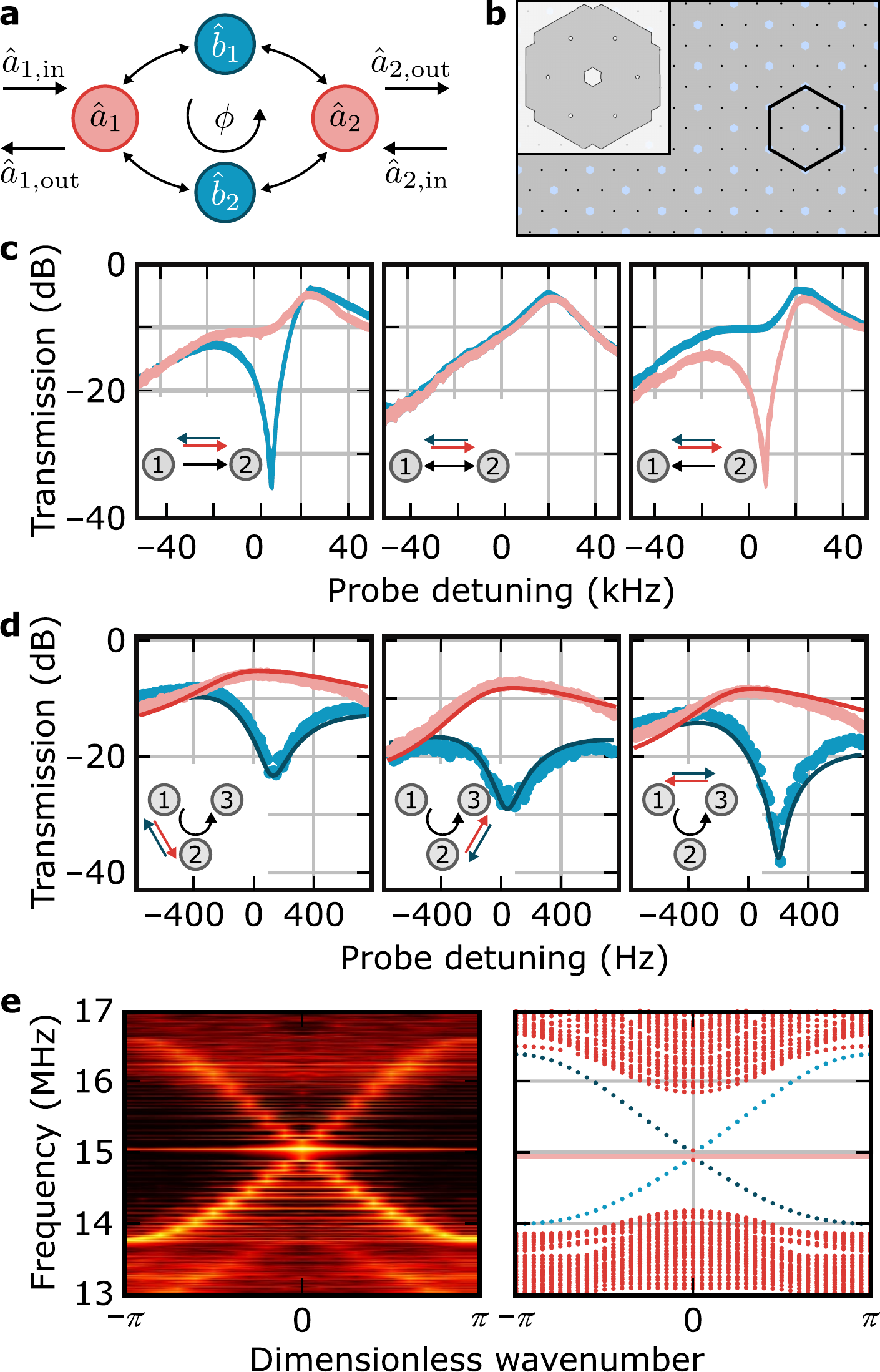}}
	\caption{\textbf{Non-reciprocal effects in optomechanical systems.} \textbf{a},~Concept underlying non-reciprocal transport in an optomechanical network.~\cite{2017NatCo...8..604B} Excitations can travel between mode $\hat{a}_1$ and $\hat{a}_2$ via the upper or lower pathways, involving mode $\hat{b}_1$ or $\hat{b}_2$, respectively; choosing the phase difference $\phi$ between the two appropriately allows for transport in one direction to be suppressed with respect to the other. \textbf{b},~A topological nanoelectromechanical metamaterial composed of a patterned SiN nanomembrane suspended over a doped silicon substrate;~\cite{2018Natur.564..229C} the hexagonal unit cell is shown in inset. \textbf{c},~Demonstration of a reconfigurable non-reciprocal optomechanical device~\cite{2017NatCo...8..604B}, showing the scattering parameters between two ports of the device; from left to right, the circuit is set up to allow transmission only from port 1 to port 2, in either direction, and only from port 2 to port 1. \textbf{d},~An experimental demonstration of a mechanical on-chip microwave circulator~\cite{2017NatCo...8..953B} set up to transmit signals from any port to the next one in a clockwise direction, but not in reverse; the figure shows the transmission between pairs of ports on the device. \textbf{e},~Frequency dispersion curves of the topological optomechanical structure shown in panel \textbf{b}~\cite{2018Natur.564..229C}, which supports topologically-protected chiral transport; the experimental measurements (left) mirror the theoretical calculations (right) closely. Figure adapted with permission from:\ panels \textbf{a} and \textbf{c}, ref.~\cite{2017NatCo...8..604B}, Springer Nature Ltd.; panel \textbf{d}, ref.~\cite{2017NatCo...8..953B}, Springer Nature Ltd.; panels \textbf{b} and \textbf{e}, ref.~\cite{2018Natur.564..229C}, Springer Nature Ltd.}
\label{Topology}
\end{figure}

\subsection{Non-reciprocal transport of photons and phonons.}
A final field of study that we choose to highlight as promising for quantum applications is that of non-reciprocal transport, one key technological application of which is in chip-scale circulators for electromagnetic radiation. In traditional large-scale optics, circulators rely on strong permanent magnets to break time-reversal symmetry. Around a decade ago it was realised that the optomechanical interaction can be exploited in place of a magnetic field~\cite{2012OExpr..20.7672H,2015PhRvX...5b1025M,2017PhRvP...7f4014M}. This opened the door to the realisation of non-reciprocal optomechanical devices~\cite{2017NatCo...8..604B,2017NatCo...8..953B,2018PhRvL.120f0601B,2017PhRvX...7c1001P,2017NatPh..13..465F,2016NatCo...713662R,2015PhRvA..91e3854X}, whose operation we illustrate in Fig.~\ref{Topology}(a–b); these devices have been used to demonstrate capabilities unique to optomechanical devices, most notably the potential for full reconfigurability~\cite{2017NatCo...8..604B,2017NatCo...8..953B} in the sense of \emph{in situ} selection of the direction of signal propagation through the device; this is demonstrated clearly in Fig.~\ref{Topology}(c–d) by observing the transmission of electromagnetic radiation between pairs of inputs and outputs.

Optomechanical platforms are uniquely suited to creating pseudo-magnetic fields~\cite{2017PNAS..114E3390B} and topological insulators~\cite{2015PhRvX...5c1011P} for light and sound. These ideas have started becoming an experimental reality only recently; one recent demonstration of topologically protected transport in an optomechanical system~\cite{2018Natur.564..229C, 2020arXiv200906174R} is shown in Fig.~\ref{Topology}(e). Non-reciprocal physics has been proposed or observed in periodic optomechanical structures, including one- and two-dimensional phononic or photonic crystals~\cite{2011PhRvL.107d3603H,2015NJPh...17b3025S,2020PhRvB.101h5108S}, arrays of mechanical resonators interacting with a single optical mode~\cite{2016OptL...41.2676G}, and other optical or mechanical platforms~\cite{2013NaPho...7.1001H,2015NatCo...6.8682M,2018PhRvB..97b0102B}.

\section{Observations, outlook, and outstanding challenges}
The interaction of electromagnetic radiation with the motion of objects has an array of implications and manifestations. The field has grown significantly since the first proof-of-principle experiments that demonstrated the use of light to dampen the motion of relatively large mechanical oscillators~\cite{2014RvMP...86.1391A}. The quests for reaching the quantum ground state of a mechanical oscillator, to put one in a non-classical state of motion, and to entangle two mechanical oscillators have all been completed. It is now time to look ahead at how optomechanical devices may be applied in forthcoming quantum technologies and as the basis for new experiments.

Looking beyond the accomplishments that we have reviewed paints an interesting picture. Fundamental exploration has always played a large role in the development of the field of optomechanics. In this spirit, we describe some interesting directions which we believe could have scientific and, eventually, technological implications: Optomechanical systems present the prospect of pushing quantum experiments to ever larger mass scales; the theoretical underpinnings of where new physics might lie – collapse models~\cite{2013RvMP...85..471B}, a resolution of the measurement problem~\cite{2004RvMP...76.1267S}, gravitational decoherence~\cite{2017CQGra..34s3002B}, to name but a few – is an open question that optomechanics is uniquely suited to explore. The dynamical Casimir effect~\cite{Moore1970,Sanz2018}, which describes the production of photons out of the vacuum due to rapidly moving boundaries, also lends itself to optomechanical realisation; recent studies encouragingly suggest the possibility of bypassing the constraint for its observation set by the demand of high mechanical frequencies~\cite{PhysRevX.8.011031}. Lastly, other interesting frontiers rely upon the unique coherence and the length-scale of mechanical excitations in optomechanical devices; active areas of investigation include using long-lived coherent phononic memories~\cite{2020NatPh..16..772W,2020Sci...370..840M,Tsaturyan2017} for quantum repeaters and interconnecting quantum processors using phononic excitations and waveguides~\cite{Fang2016,2018PhRvL.121d0501P,2021arXiv210806248Z}.

We expect that the research landscape will increasingly focus on refining the applications of optomechanical devices in quantum technologies, on the one hand, and on understanding fundamental phenomena like the ones above, on the other. Cross-fertilisation between these two branches will likely remain a hallmark of the field, bringing new insight and advances to materials science, nano- and micro-structure fabrication, and fundamental quantum science to investigate, among other topics, the emergence of new physics.

\begin{acknowledgments}
S.B.\ acknowledges funding by the Natural Sciences and Engineering Research Council of Canada (NSERC) through its Discovery Grant, funding and advisory support provided by Alberta Innovates through the Accelerating Innovations into CarE (AICE) – Concepts Program, and support from Alberta Innovates and NSERC through Advance Grant. A.X.\ acknowledges funding by the European Union’s Horizon 2020 research and innovation programme under grant agreement no.\ 732894 (FET Proactive HOT) and by the Julian Schwinger Foundation project grant no.\ JSF-16-03-0000 (TOM). S.G.\ is supported by the European Research Council (ERC StG Strong-Q, 676842; and ERC CoG Q-ECHOS, 101001005), by the Netherlands Organization for Scientific Research (NWO/OCW) as part of the Frontiers of Nanoscience program, as well as through Vidi (680-47-541/994) and Vrij Programma (680-92-18-04) grants. M.P.\ is supported by the H2020\-/FETOPEN\-/2018\-/2020 project TEQ (grant 766900), the DfE-SFI Investigator Programme (grant 15/IA/2864), COST Action CA15220, the Royal Society Wolfson Research Fellowship (RSWF\textbackslash R3\textbackslash183013), the Royal Society International Exchanges Programme (IEC\textbackslash R2\textbackslash192220), the Leverhulme Trust Research Project Grant (grant RGP\-/2018\-/266), the UK EPSRC (project QuamNESS, grant EP/T028106/1), and the CNR/RS (London) project ``Testing fundamental theories with ultracold atoms.'' C.A.R.\ acknowledges funding by the US National Science Foundation under Grant No.\ 1125844 and a Cottrell FRED Award from the Research Corporation for Science Advancement under grant 27321. E.M.W.\ acknowledges funding by the European Union’s Horizon 2020 research and innovation program under grant agreement No 732894 (FET Proactive HOT), the German Federal Ministry of Education and Research (contract no.\ 13N14777) within the European QuantERA co-fund project QuaSeRT, and via project QT-6 SPOC of the Baden-W\"urttemberg Foundation.
\end{acknowledgments}

\end{document}